\documentclass[twocolumn,amsmath,amssymb,pre]{revtex4-1}

\usepackage{graphicx}
\usepackage{color}
\usepackage{dcolumn}
\usepackage{amsmath}
\usepackage{CJK}
\usepackage{subfigure}
\usepackage{sidecap}

\usepackage{mathrsfs}
\usepackage{amsfonts}
\usepackage{indentfirst}
\usepackage{bm,bbm}

\usepackage[colorlinks,citecolor=blue]{hyperref}  

\usepackage[T1]{fontenc}

\newcommand{\be}{\begin{equation}}
\newcommand{\ee}{\end{equation}}
\newcommand{\bey}{\begin{eqnarray}}
\newcommand{\eey}{\end{eqnarray}}
\newcommand{\bw}{\begin{widetext}}
\newcommand{\ew}{\end{widetext}}

\newcommand{\ra}{\rangle}
\newcommand{\la}{\langle}

\newcommand{\ba}{\begin{array}}
\newcommand{\ea}{\end{array}}
\newcommand{\bi}{\begin{itemize}}
\newcommand{\ei}{\end{itemize}}
\newcommand{\bem}{\begin{enumerate}}
\newcommand{\eem}{\end{enumerate}}

\begin{document}

\title{Performance of quantum heat engines under the influence of long-range interactions}

\author{Qian Wang\footnote{Electronic address: qwang@zjnu.edu.cn}}

\affiliation{Department of Physics, Zhejiang Normal University, Jinhua 321004, China}
\affiliation{CAMTP-Center for Applied Mathematics and Theoretical Physics, University of Maribor, Mladinska 3, SI-2000
Maribor, Slovenia}


\begin{abstract}

We examine a quantum heat engine with an interacting many-body working medium consisting of
the long-range Kitaev chain to explore the role of long-range interactions in the performance of the quantum engine.
By analytically studying two types of thermodynamic cycles, namely, the Otto cycle and Stirling cycle, we demonstrate that
the work output and efficiency of a long-range interacting heat engine can be boosted by the long-range interactions,
in comparison to the short-range counterpart. 
We further show that in the Otto cycle there exists an optimal condition for which the 
maximum enhancement in work output and efficiency can be achieved simultaneously by the long-range interactions. 
But, for the Stirling cycle, the condition which can give the maximum enhancement in work output does not
lead to the maximum enhancement in efficiency. 
We also investigate how the parameter regimes under which the engine performance is enhanced 
by the long-range interactions evolve with a decrease in the range of interactions.

\end{abstract}

\date{\today}

\maketitle

\section{introduction}

Since the seminal work of Scovil and Schulz-DuBois \cite{Scovil1959}, the concept of quantum heat engines 
has attracted lots of attention \cite{Alicki1979,Geva1992,Scully2003,Kieu2004,Quan2005,Quan2006,Henrich2007a,Quan2007,
Quan2009,Kosloff2014}.  
In particular, in the last few decades, spurred by experimental and theoretical progress in the field of quantum 
thermodynamics (see, e.~g., Refs.~\cite{Binder2018,Sai2016,Deffner2019} and references therein), a tremendous 
amount of effort has been devoted to studies of quantum heat engines,
both experimentally \cite{Koski2014,Rossnagel2016,Klaers2017,Zou2017,Klatzow2019,Assis2019,Lindenfels2019,Peterson2019} 
and theoretically \cite{Abah2014,Gelbwaser2015,Uzdin2015,Abiuso2020,Kosloff2017n,Deffner2019}.
As the quantum versions of classical engines, quantum heat engines exploit quantum systems as 
their working medium to extract work in quantum thermodynamic cycles \cite{Kieu2004,Quan2007}.  
A wide range of quantum systems have been used to devise quantum heat engines, including 
harmonic oscillators \cite{Kosloff2017n,Insinga2016,Reid2017},
spin systems \cite{Thomas2011,Altintas2015,Campisi2015,Marchegiani2016,Barrios2017,Hewgill2018}, 
photons \cite{Scully2003,Dillenschneider2009}, 
optomechanical systems \cite{Keye2014}, Dirac particles \cite{Munoz2012,Pena2016}, 
and quantum dots \cite{Josefsson2018,Pena2019,Pena2020}.
A few works have also considered quantum engines with the heat reservoirs replaced by quantum systems 
\cite{Abah2014,ZhangX2014,Niedenzu2018}. 

Numerous studies in quantum heat engines are focused on possible enhancement in work output and efficiency via
the utilization of quantum properties in various working mediums 
\cite{Hardal2015,Hewgill2018,Horne2018,Watanabe2017} 
or heat reservoirs \cite{Scully2003,Abah2014,Klaers2017,Niedenzu2018}.
Other efforts are towards understanding the fundamental differences between classical and 
quantum heat engines \cite{Quan2007,Gardas2015,Friedenberger2017},
finite time cycles \cite{Esposito2010,Feldmann2012,WuF2014,ZhengY2016,Cavina2017,Wiedmann2019}, 
and the applications of shortcuts to adiabaticity  
\cite{DengJ2013,ACampo2014,LiT2018,Abah2018,Abah2018,Abah2019,Cakmak2019,Funo2019,Hartmann2019}.
The effects of quantum statistics on the performance of quantum heat engines have 
also been investigated \cite{ZhengYP2015,Watanabe2019}. 
Very recently, the results in Ref.~\cite{Myers2020} further demonstrated that wave-function symmetry 
has nontrivial impacts on the performance of quantum heat engines.  

Most of the aforementioned works are limited to the single or few-particle working mediums.
However, with the aim to scale up quantum heat engines and related thermodynamic devices, 
it is necessary for us to consider many-body quantum heat engines.
Several recent studies have reported that engine performance can be enhanced by   
various quantum many-body effects, such as 
quantum phase transitions \cite{Campisi2016,MaY2017,Fadaie2018,Kloc2019,Revathy2020} and 
many-body localization \cite{Halpern2019}.
Importantly, the significant role played by interparticle interactions 
in quantum many-body heat engines has been revealed in Refs.~\cite{Jaramillo2016,Beau2016,Chen2019}.

On the other hand, the recent experimental realization of quantum many-body systems 
with tunable long-range interactions \cite{Jurcevic2014,Richerme2014}
has triggered a surge of interest in the properties of quantum systems 
with long-range interactions 
\cite{Santos2016,Halimeh2017,GongZ2017,Jaschke2017,Silva2018,LiuF2019,Lerose2019,Piccitto2019,
Puebla2019,Piotr2019}.
In these systems, the interaction strength between two particles with distance $r$ 
usually decays as $1/r^\alpha$ with $\alpha\geq0$. 
Due to the long-range interactions, long-range interacting systems can host novel features 
that are not observed in their short-range counterparts, 
such as the anomalous dynamical phase \cite{Homrighausen2017}, the quantum time crystal without an external driver 
\cite{Kozin2019}, and the breakdown of locality (see Refs.~\cite{LiuF2019,Schachenmayer2013,Buyskikh2016} and references therein).
Given the fact that the long-range interactions are very relevant in experimental platforms that include 
Rydberg atoms \cite{Saffman2010}, trapped ions \cite{Jurcevic2014,Richerme2014,Britton2012,Islam2013},
polar molecules \cite{Bohn2017}, and magnetic atoms \cite{Lahaye2009},
it is thus interesting to see whether the long-range interactions 
can be used to enhance engine performance.  

In this work we explore the enhancement effect of long-range interactions 
on the performance of a quantum heat engine which
employs the long-range Kitaev chain as its working medium.
The long-range Kitaev chain can be considered an extension of the well-known Kitaev chain \cite{Kitaev2001} 
with a long-range superconducting pairing term and has been used as a prototypical model in the studies
of long-range interacting systems \cite{Vodola2014,Vodola2015phD,Maity2019}.
By exploiting the integrability of the long-range Kitaev chain, we are able to obtain the explicit expressions for 
the work output and efficiency of the quantum engine. 
We will consider two types of thermodynamic cycles, namely, the Otto cycle and Stirling cycle, respectively. 
For both types of cycles, we find that the long-range interactions can display a notable improvement in engine performance
in comparison to the case of short-range interactions. 
We further demonstrate how the enhancement regions in the cycle parameter space evolve with the range of interactions. 
Here we stress that we consider only the quasistatic cycles, leaving the cycles that operate 
in finite time as an interesting topic of future study.
Moreover, we focus on the engine performance in terms of work output and efficiency.

The remainder of this article is organized as follows. 
In Sec.~\ref{Model} we introduce the long-range interacting Kitaev chain and review briefly some of its basic features; 
we also specify the exact expressions for the thermodynamic quantities of the long-range Kitaev chain in this section.
In Sec.~\ref{QHEs} we describe two types of thermodynamic cycles (i. e., the Otto cycle and the Stirling cycle) 
in which our quantum heat engine operates and we extract the analytical expressions for work output and the efficiency 
of these cycles, respectively.  
Then we present our numerical results and a discussion in Sec.~\ref {NRD}. 
We finally conclude our study in Sec.~\ref{ConS}.

\section{The long-range Kitaev Chain} \label{Model}

We consider the one-dimensional Kitaev model with long-range pairing interactions. Its Hamiltonian reads
\cite{Vodola2014,Pezze2017,Maity2019,Vodola2015phD,Dutta2017}
\begin{align} \label{LRKH}
 H=&-J\sum_{j=1}^L(c_j^\dag c_{j+1}+c_{j+1}^\dag c_j)-\mu\sum_{j=1}^L\left(c_j^\dag c_j-\frac{1}{2}\right) \notag \\
 &+\frac{\Delta}{2}\sum_{j=1}^L\sum_{\ell=1}^{L-1}\frac{1}{d_\ell^\alpha}\left(c_j c_{j+\ell}+c_{j+\ell}^\dag c_j^\dag \right).
\end{align}
Here, $c_j^\dag (c_j)$ are the fermionic creation (annihilation) operators on the $j$th site, $L$ denotes the 
size of the model, and $\mu$ is the on-site chemical potential. $J$ represents the hopping 
strength of the fermion between nearest neighbor sites, while $\Delta$ is the strength of the fermion pairing interactions. 
The power-law decaying
pairing term is characterized by the exponent $\alpha$. $d_\ell=\mathrm{min}(\ell,L-\ell)$ 
is the effective distance between the $i$th and $(i+\ell)$th sites. 
In our study, we consider a close chain with 
an antiperiodic boundary condition $c_j=-c_{j+L}$ and assume the total number of sites $L$ is even.
Throughout this work, we set $\hbar=1$.

  \begin{figure}
    \includegraphics[width=\columnwidth]{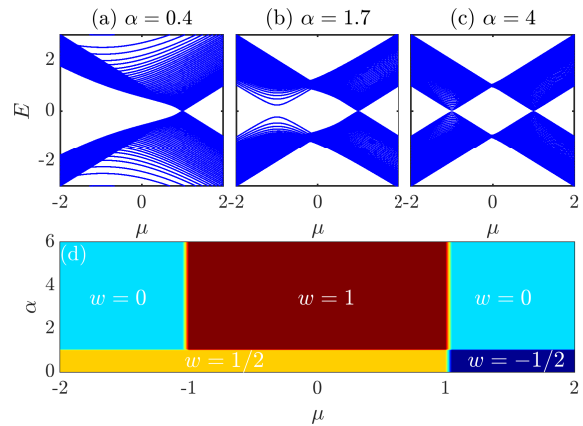}
  \caption{Energy spectrum as a function of $\mu$ of the long-range Kitaev chain (\ref{LRKH}) with antiperiodic 
  boundary condition for (a) $\alpha=0.4$, (b) $\alpha=1.7$, and (c) $\alpha=4$. 
  The system size is $L=200$, $J=\Delta=1$, and $\hbar=1$. 
  (d) Schematic phase diagram of the long-range Kitaev chain in the $\mu-\alpha$ plan.
  Different phases are indicated by different colors with corresponding winding number $w$.
  All quantities are unitless.}
  \label{phaseD}
 \end{figure}

In the short-range limit $\alpha\to\infty$, the Hamiltonian in Eq.~(\ref{LRKH}) reduces to the well-known 
short-range Kitaev chain \cite{Kitaev2001},
which can be mapped to the quantum transverse field Ising model via the Jordan-Wigner transformation 
\cite{Lieb1961}. 
In this case, by analytically diagonalizing the Hamiltonian, one can find that the model undergoes the topological phase
transitions at the critical points $\mu/J=\pm1$. 
For finite $\alpha$, the above-mentioned mapping does not exist anymore. However, the quadratic form of the 
Hamiltonian (\ref{LRKH}) implies that it can still be exactly diagonalized for any finite $\alpha$.

To this end, we first recast the Hamiltonian (\ref{LRKH}) in the momentum space
via the Fourier transform $c_j=L^{-1/2}\sum_kc_ke^{-ikj}$, where due to an antiperiodic boundary condition, 
the lattice momentum $k=\pm\pi(2n-1)/L$ with $n=1,2,\ldots,L/2$. 
Then the Hamiltonian takes the following block diagonal form \cite{Maity2019,Pezze2017,Dutta2017}
\be\label{blockDH}
  H=\sum_k\mathcal{C}^\dag_k\mathcal{H}_k\mathcal{C}_k,
\ee
where $\mathcal{C}_k^\dag=(c_k^\dag, c_{-k})$ and
\be
  \mathcal{H}_k=
  \begin{bmatrix}
    -(J\cos k+\mu)  & i \Delta f_\alpha(k)/2 \\
    -i \Delta f_\alpha(k)/2 & J\cos k+\mu
  \end{bmatrix},
\ee
with the function $f_\alpha(k)=\sum_{\ell=1}^{L-1}\sin(k\ell)/d_\ell^\alpha$.

The Hamiltonian (\ref{blockDH}) can be diagonalized through the Bogoliubov transformation 
\be
 \begin{pmatrix}
   c_k \\
   c_{-k}^\dag
 \end{pmatrix}
 =
 \begin{pmatrix}
   \cos\theta_k & i\sin\theta_k \\
   i\sin\theta_k & \cos\theta_k
 \end{pmatrix}
 \begin{pmatrix}
   d_k \\
   d_{-k}^\dag
 \end{pmatrix}.
\ee
Here, the Bogoliubov angle $\theta_k$ is defined as
\be\label{BGA}
  \tan(2\theta_k)=-\frac{\Delta f_\alpha(k)/2}{J\cos k+\mu}.
\ee
The Hamiltonian (\ref{LRKH}) is finally diagonalized as
\be\label{DGH}
  H=\sum_k\varepsilon_k\left(d^\dag_k d_k-\frac{1}{2}\right),
\ee
where $\varepsilon_k$ is the quasiparticle energy and is given by
\be\label{quasiE}
 \varepsilon_k=\sqrt{(J\cos k+\mu)^2+[\Delta f_\alpha(k)/2]^2}.
\ee
For the short-range limit $\alpha\to\infty$, we have $f_\infty(k)=2\sin k$ \cite{Maity2019,Pezze2017}. 
Then the quasiparticle energy $\varepsilon_k$ takes the usual short-range form \cite{Kitaev2001}. 
However, in the thermodynamic limit $L\to\infty$, the function $f_\alpha(k)$ becomes 
$f_\alpha^\infty(k)=(-i/2)\left[\mathbf{Li}_\alpha(e^{ik})-\mathbf{Li}_\alpha(e^{-ik})\right]$,
with $\mathbf{Li}_\alpha(x)=\sum_{\ell=1}^\infty(x^\ell/\ell^\alpha)$ being the $\alpha$th-order polylogarithm function of $x$
\cite{Pezze2017,Dutta2017,Maity2019,Vodola2015phD}.
It is known that for $\alpha>1$, $f_\alpha^\infty(k)$ vanishes at $k=0$ and $k=\pi$, while when $\alpha<1$, 
$f_\alpha^\infty(k)=0$ occurs only at $k=\pi$ \cite{Pezze2017,Maity2019,Dutta2017}.
Thus for finite $\alpha$, the spectrum is gapless at $\mu/J=\pm1$ when $\alpha>1$, whereas once $\alpha<1$, 
we have $\varepsilon_k=0$
only at $\mu/J=1$. For brevity but without loss of generality, we set $J=\Delta=1$ in the following part of our study.

  \begin{figure}
    \includegraphics[width=\columnwidth]{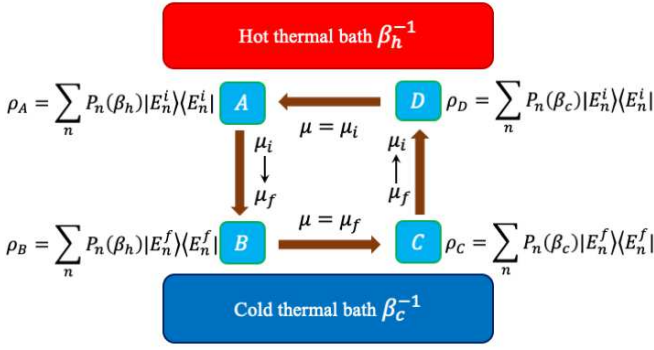}
  \caption{Schematic diagram of the quantum Otto cycle studied in this work.
  The thermodynamic cycle operats between a hot bath at temperature $\beta_h^{-1}$ 
  and a cold bath at temperature $\beta_c^{-1}$.
  It consists of two adiabatic ($A\to B$ and $C\to D$) and two isochoric ($B\to C$ and $D\to A$) strokes.
  $\mu_i$ and $\mu_f$ are the chemical potential in the two isochoric strokes.
  $\rho_\gamma$ ($\gamma=A, B, C, D$) denotes the state of the working medium at each stage of the cycle.
  At stages $A$ and $C$ the working medium is in thermal equilibrium with the hot and cold baths, respectively.}
  \label{OttoCy}
 \end{figure}

For the long-range Kitaev chain with antiperiodic boundary conditions,  
the energy spectrum with respect to the different chemical potential $\mu$ for several values of $\alpha$
is plotted in panels (a)--(c) of Fig.~\ref{phaseD}. 
Clearly, regardless of $\alpha$, the energy gap is always close to zero at $\mu=1$. 
In contrast, the gap at $\mu=-1$ is increased with decreasing $\alpha$.
In the thermodynamic limit $L\to\infty$, one can expect that the energy gap closes at $\mu=\pm1$ 
for $\alpha>1$, while it closes only at $\mu=1$ when $\alpha<1$. 
The close of the energy gap in the spectrum corresponds to the transitions between different topological phases,
which are characterized by different winding numbers.
The winding number is defined as $w=(1/2\pi)\int_{-\pi}^\pi(\partial_k\theta_k)dk$, with $\theta_k$ 
the Bogoliubov angle given by Eq.~(\ref{BGA}). 
Figure~\ref{phaseD}(d) shows the schematic phase diagram of the long-range Kitaev chain in the $\mu-\alpha$ plan.
The topological phases with different winding numbers are discriminated by different color regions.
Depending upon the values of $\mu$ and $\alpha$, distinct topological phases with $w=\pm1/2,0,1$ are displayed in the model.
It is worth mentioning that the effects of the topological phase transition on the work and efficiency of the 
quantum heat engine have been explored in Refs.~\cite{Fadaie2018,Yunt2019}.
In our study we mainly focus on the influences of the long-range interactions 
on the performance of quantum heat engines.
Moreover,  in order to avoid the divergence in thermodynamic limit for $0\leq\alpha\leq1$, we restrict our study to
$\alpha>1$, as is done in other studies of long-range interacting systems \cite{LiuF2019,Dutta2017}.

 \begin{figure}
    \includegraphics[width=\columnwidth]{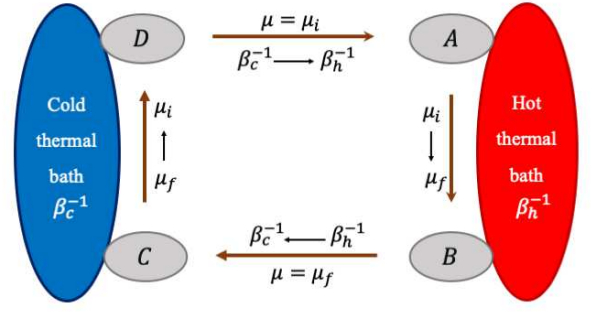}
  \caption{Graphic sketch of the quantum Stirling cycle operating between 
  two thermal baths (hot and cold) at temperatures $\beta_h^{-1}$ and $\beta_{c}^{-1}$, respectively.
  The cycle has two isothermal ($A\to B$ and $C\to D$) and two isochemical ($B\to C$ and $D\to A$) potential processes.
  $\mu_i$ and $\mu_f$ are the values of the chemical potential in the two isochemical potential processes.}
  \label{StirlingCy}
 \end{figure}

\subsection*{Expressions of the thermodynamic quantities}

Considering the model is in an equilibrium state at temperature $\beta^{-1}$, here and henceforth we set $k_B=1$.
The state of the model is, therefore,
the Gibbs state $\rho=e^{-\beta H}/\mathcal{Z}$, where $\mathcal{Z}=\mathrm{tr}(e^{-\beta H})$ is the partition function. 
This, of course, assumes that the energy of the system is the only conserved quantity and the thermodynamics of the system
is calculated by the standard Gibbs measure. 
From the diagonal form of the Hamiltonian in Eq.~(\ref{DGH}), the partition function can be written as
\be
   \mathcal{Z}=2^L\prod_{k>0}\cosh^2\left(\frac{\beta\varepsilon_k}{2}\right).
\ee
Then the internal energy $U$ and the free energy $F$ for the long-range Kitaev chain are respectively given by
\begin{align}
  &U=\mathrm{tr}(\rho H)=-\frac{\partial\ln\mathcal{Z}}{\partial\beta}
       =-\sum_{k>0}\varepsilon_k\tanh\left(\frac{\beta\varepsilon_k}{2}\right), \label{UEG}\\
  &F=-\frac{\ln\mathcal{Z}}{\beta}=-\frac{1}{\beta}\left[L\ln2+\sum_{k>0}2\ln\cosh\left(\frac{\beta\varepsilon_k}{2}\right)\right].
\end{align}
With the density of state $\rho$, one can find the entropy $S=-\mathrm{Tr}[\rho\ln\rho]$ has the from
\begin{align} \label{ENP}
  S=L\ln2+\sum_{k>0}\left[2\ln\cosh\left(\frac{\beta\varepsilon_k}{2}\right)
       -\beta\varepsilon_k\tanh\left(\frac{\beta\varepsilon_k}{2}\right)\right].
\end{align}
Armed with these results, we turn to investigating the effects of the interacting range on the performance of the
quantum heat engine.

\section{Thermodynamic cycles} \label{QHEs}

In our study, we analyze the performance of a quantum heat engine with the 
long-range Kitaev chain in Eq.~(\ref{LRKH}) as its working medium. 
We focus on two types of thermodynamic cycles, namely, the Otto cycle and Stirling cycle.

  \begin{figure}
    \includegraphics[width=\columnwidth]{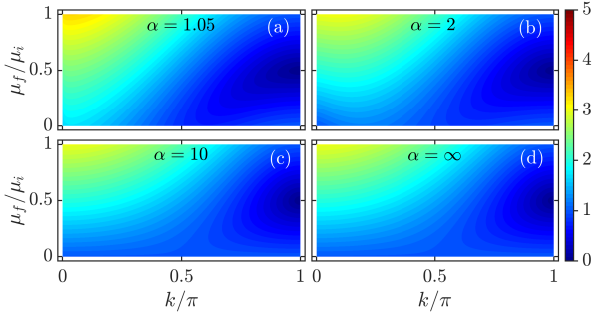}
  \caption{Quasiparticle energy $\varepsilon_k$ (\ref{quasiE}) as a function of $k$ (rescaled by $\pi$) 
  and $\mu_f$ (rescaled by $\mu_i$) for different ranges of 
  interactions: (a) $\alpha=1.05$, (b) $\alpha=2$, (c) $\alpha=10$, and (d) $\alpha=\infty$.
  Parameters are: $\hbar=k_B=1$, $J=\Delta=1$, $L=2000$, and $\mu_i=2$.}
  \label{EngV}
 \end{figure}

\subsection{Quantum Otto cycle} \label{QOC}

As a widely used cycle in both theoretical and experimental studies of quantum heat engines,
the quantum Otto cycle consists of two isochoric and two adiabatic strokes \cite{Henrich2007a,Quan2007}, 
see Fig.~\ref{OttoCy}. 
In the isochoric branches, the working medium with fixed Hamiltonian is coupled 
to the hot and cold baths and thermalized with the baths. 
For the adiabatic strokes of the cycle, the working medium is detached from the hot and cold baths, and undergoes
a unitary evolution by changing the control parameter in the Hamiltonian in an infinitely slow way so that
the quantum adiabatic theorem holds.    
The details of the four strokes are described as follows.

{\it (a) Adiabatic stroke $A\to B$ (expansion process).} 
The working medium is initially in the thermal state $\rho_A=\sum_n P_n(\beta_h)|E_n^i\ra\la E_n^i|$ at temperature 
$\beta_h^{-1}$ and Hamiltonian $H_i$, which gives by Eq.~(\ref{LRKH}) with $\mu=\mu_i$.
Here, $|E_n^i\ra$ is the $n$th eigenstate of the Hamiltonian $H_i$ with corresponding eigenvalue $E_n^i$, and 
$P_n(\beta_h)=e^{-\beta_h E_n^i}/\mathcal{Z}_h$ with 
$\mathcal{Z}_h=\sum_n e^{-\beta_h E_n^i}$. 
Then we decouple the working medium from the thermal bath and slowly vary the chemical potential from
$\mu_i$ to $\mu_f$ ($\mu_f\leq\mu_i$) such that the quantum adiabatic theorem holds in this process.
Hence, during the process the occupation probability of each eigenstate of the working medium remains the same; 
no heat is transferred. 
However, the change in the chemical potential $\mu$ results in the work being done in this process.  
At the end of the stroke, the working medium 
reaches the state $\rho_B=\sum_nP_n(\beta_h)|E_n^f\ra\la E_n^f|$, 
where $\{|E_n^f\ra\}$ are the eigenstates of $H_f$, which is the Hamiltonian (\ref{LRKH}) with $\mu=\mu_f$.

{\it (b) Isochoric stroke $B\to C$ (cooling process).} In this stroke, the chemical potential is fixed at $\mu=\mu_f$ and  
the working medium is brought in contact with the cold bath at temperature $\beta_c^{-1}$
until it attains the thermal equilibrium with the cold bath.
The state of the working medium at the end of this process is, therefore, given by
$\rho_C=\sum_nP_n(\beta_c)|E_n^f\ra\la E_n^f|$,
with $P_n(\beta_c)=e^{-\beta_c E_n^f}/\mathcal{Z}_c$ and $\mathcal{Z}_c=\sum_ne^{-\beta_c E_n^f}$.
Because the chemical potential is fixed, no work has been done in this process,
and the heat $Q_O^c$ is ejected from the working medium to the cold bath.  

{\it (c) Adiabatic stroke $C\to D$ (compression process).} 
The working medium is detached from the cold bath and the chemical potential is
adiabatically driven back from $\mu_f$ to $\mu_i$. The occupation probability of each energy levels stays at the 
values $P_n(\beta_c)$. The work is done in this process but no heat transfer. 
At the end of this stroke, the state of the working medium reads
$\rho_D=\sum_n P_n(\beta_c)|E_n^i\ra\la E_n^i|$. 

{\it (d) Isochoric stroke $D\to A$ (heating process).} 
This is an isochoric heating process in which the working medium, with chemical potential $\mu=\mu_i$ 
and Hamiltonian $H=H_i$, is attached with the hot bath at temperature $\beta^{-1}_h$, 
relaxing to the thermal state $\rho_A$. 
In this process, the fixed chemical potential means that the energy levels are unchanged, no work is done.
But the change in the occupation probability of each eigenstate causes the working medium 
to absorb heat $Q_O^h$ from the hot bath. 

As no work is done in the two isochoric strokes of the cycle, the heat transfer between 
the working medium and the heat bath
in the isochoric heating and cooling strokes is equal to the variation of internal energy during the processes.
Therefore the heat $Q_O^h$ injected into the working medium in the heating stroke and heat $Q_O^c$ ejected
in the cooling stroke are calculated as
$Q_O^h=\mathrm{Tr}[H_i(\rho_A-\rho_D)], Q_O^c=\mathrm{Tr}[H_f(\rho_C-\rho_B)]$ 
\cite{Quan2007,Henrich2007a,Thomas2011}.
By using Eq.~(\ref{UEG}), one can explicitly write $Q_O^h$ and $Q_O^c$ as follows:
\begin{align}
  Q_O^h&=\sum_{k>0}\varepsilon_k^i\left[\tanh\left(\frac{\beta_c\varepsilon_k^f}{2}\right)-
                 \tanh\left(\frac{\beta_h\varepsilon_k^i}{2}\right)\right], \\
  Q_O^c&=\sum_{k>0}\varepsilon_k^f\left[\tanh\left(\frac{\beta_h\varepsilon_k^i}{2}\right)-
                 \tanh\left(\frac{\beta_c\varepsilon_k^f}{2}\right)\right], 
\end{align}  
where $\varepsilon_k^i$ and $\varepsilon_k^f$ are given by Eq.~(\ref{quasiE}), where 
$\mu$ is replaced by $\mu_i$ and $\mu_f$, respectively. 

The work is done only in the quantum adiabatic strokes of the cycle. From the first law of thermodynamics,
the amount of net work performed by the Otto cycle is given by
\begin{align} \label{NetW}
   W_O&=Q_O^h+Q_O^c \notag \\
          &=\sum_{k>0}
           \left(\varepsilon_k^i-\varepsilon_k^f\right)\left[\tanh\left(\frac{\beta_c\varepsilon_k^f}{2}\right)-
                 \tanh\left(\frac{\beta_h\varepsilon_k^i}{2}\right)\right].
\end{align}
Note that in order to make sure the cycle works as an engine,
we must have $W_O>0$, $Q_O^h>-Q_O^c>0$.  
Then the efficiency of the engine is given by 
\be
 \eta_O=\frac{W_O}{Q_O^h}.
\ee
It is known that the efficiency of the quantum Otto cycle is bounded by 
the Carnot efficiency $\eta_c=1-\beta_h/\beta_c$ \cite{Quan2007}.

  \begin{figure}
    \includegraphics[width=\columnwidth]{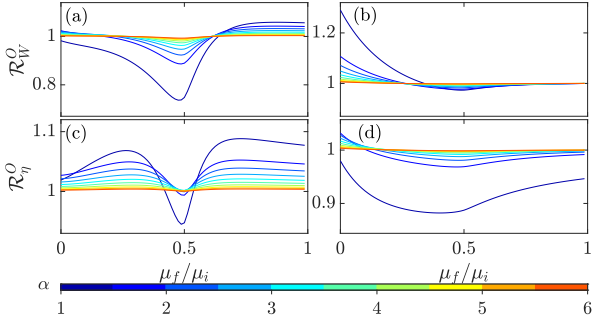}
  \caption{Left column: Work output ratio $\mathcal{R}_{W}^O$ (a) and the efficiency ratio $\mathcal{R}_\eta^O$ (c)
   of Otto cycle, see Eq.~(\ref{DfR}), as a function of $\mu_f/\mu_i$ for $\alpha\in[1.05,6]$ at $\beta_c=5$.
  Right column: $\mathcal{R}_W^O$ (b) and  $\mathcal{R}_{\eta}^O$ (d)
  as a function of $\mu_f/\mu_i$ for $\alpha\in[1.05,6]$ at $\beta_c=0.05$.
  Other parameters are: $\hbar=k_B=1$, $\beta_h/\beta_c=0.2$, $\mu_i=2$, $J=\Delta=1$, and $L=2000$.}
  \label{OtRv}
 \end{figure}

\subsection{Quantum Stirling cycle} \label{QSC}

In this section we consider the quantum Stirling cycle, which has been used to explore
the quantum criticality impacts on the performance of quantum heat engines \cite{MaY2017,Fadaie2018}.  
The Stirling cycle is composed of two isothermal 
and two isochemical potential (i.e., fixed value of chemical potential) processes. 
As shown in Fig.~\ref{StirlingCy}, the details of the cycle are given as follows.

{\it (a) Isothermal process ($A\to B$).} 
In this process the working medium is kept contact with the hot bath at temperature $\beta_h^{-1}$.
The chemical potential is changed from $\mu_i$ to $\mu_f$, 
and the amount of heat, denoted by $Q_{\mathrm{I}}$, absorbed by the working medium from the hot bath 
is $Q_{\mathrm{I}}=(S_f-S_i)/\beta_h$, where $S_i$ ($S_f$) is the entropy of the working medium with
$\mu=\mu_i$ ($\mu=\mu_f$) at temperature $\beta_h^{-1}$. 
From the expression of entropy in Eq.~(\ref{ENP}), $Q_{\mathrm{I}}$ can be calculated as
\begin{align}
  Q_{\mathrm{I}}=&\sum_{k>0}\left\{\frac{2}{\beta_h}\left[\ln\cosh\left(\frac{\beta_h\varepsilon_k^f}{2}\right)-
   \ln\cosh\left(\frac{\beta_h\varepsilon_k^i}{2}\right)\right]\right. \notag \\
   &\left.-\left[\varepsilon_k^f\tanh\left(\frac{\beta_h\varepsilon_k^f}{2}\right)
   -\varepsilon_k^i\tanh\left(\frac{\beta_h\varepsilon_k^i}{2}\right)\right]\right\}.
\end{align}
Here $\varepsilon_k^i$ ($\varepsilon_k^f$) is defined by Eq.~(\ref{quasiE}) with $\mu=\mu_i$ ($\mu=\mu_f$).

{\it (b) Isochemical potential process ($B\to C$):} 
The chemical potential is fixed at $\mu=\mu_f$. The temperature of the working medium 
is decreased from $\beta_h^{-1}$ to $\beta_c^{-1}$. 
Because the chemical potential is fixed, the energy levels of the working medium remain invariant. 
Hence, no work is done during 
this process, but heat $Q_{\mathrm{II}}=U_f(\beta_c)-U_f(\beta_h)<0$ is released to the reservoir.
Here, $U_f(\beta_c)$ [$U_f(\beta_h)]$ is the internal energy of the working medium 
at temperature $\beta_c^{-1}$ ($\beta_h^{-1}$) with $\mu=\mu_f$.
By employing Eq.~(\ref{UEG}), $Q_{\mathrm{II}}$ can be expressed as
\be
  Q_{\mathrm{II}}=\sum_{k>0}\varepsilon_k^f\left[\tanh\left(\frac{\beta_h\varepsilon_k^f}{2}\right)-
         \tanh\left(\frac{\beta_c\varepsilon_k^f}{2}\right)\right].
\ee

{\it (c) Isothermal process ($C\to D$):} 
The working medium undergoes another isothermal process which restores the chemical potential 
to the value $\mu_i$ from $\mu_f$ while keeping the working medium in contact 
with the cold bath at temperature $\beta_c^{-1}$. In this process, the working medium ejects heat 
$Q_{\mathrm{III}}=(S_i-S_f)/\beta_c<0$, which can be explicitly written as
\begin{align}
  Q_{\mathrm{III}}=&\sum_{k>0}\left\{\frac{2}{\beta_c}\left[\ln\cosh\left(\frac{\beta_c\varepsilon_k^i}{2}\right)-
   \ln\cosh\left(\frac{\beta_c\varepsilon_k^f}{2}\right)\right]\right. \notag \\
   &\left.-\left[\varepsilon_k^i\tanh\left(\frac{\beta_c\varepsilon_k^i}{2}\right)
   -\varepsilon_k^f\tanh\left(\frac{\beta_c\varepsilon_k^f}{2}\right)\right]\right\}.
\end{align}

{\it (d) Isochemical potential process ($D\to A$):}  
As the reverse of the process (II), the last process of the cycle is operated at 
fixed chemical potential $\mu_i$ and the temperature of the working medium is 
increased from $\beta_c^{-1}$ to $\beta_h^{-1}$.
During this process, the heat absorbed by the working medium is given by
$Q_{\mathrm{IV}}=U_i(\beta_h)-U_i(\beta_c)$, where $U_i(\beta_h)$ [$U_i(\beta_c)$] denotes the internal energy of
the working medium at temperature $\beta_h^{-1}$ ($\beta_c^{-1}$) with $\mu=\mu_i$.
Using Eq.~(\ref{UEG}), $Q_\mathrm{IV}$ takes the form
\be
  Q_\mathrm{IV}=\sum_{k>0}\varepsilon_k^i\left[\tanh\left(\frac{\beta_c\varepsilon_k^i}{2}\right)-
         \tanh\left(\frac{\beta_h\varepsilon_k^i}{2}\right)\right].
\ee

 \begin{figure}
  \includegraphics[width=\columnwidth]{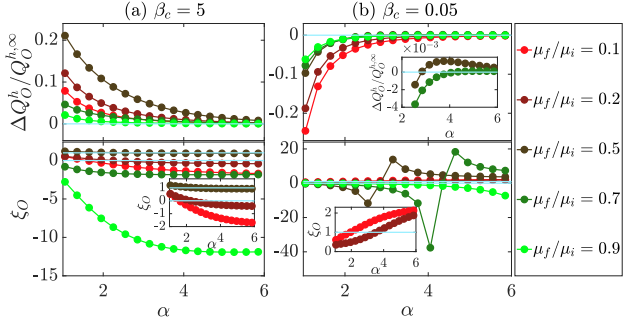}
  \caption{Column (a): $\Delta Q_O^h/Q_O^{h,\infty}$ (upper panel) and $\xi_O$
  (bottom panel and zoomed-in inset) as a function of $\alpha$ for different values of $\mu_f/\mu_i$ at $\beta_c=5$.  
  Column (b): The same as in column (a) applied to the high-temperature case with $\beta_c=0.05$. 
  The solid horizontal lines in the upper panels and associated inset indicate $\Delta Q_O^h=0$.
  The horizontal lines in the bottom panels and associated insets indicate $\xi_O=1$ and $\xi_O=0$.  
  The remaining parameters are: $\hbar=k_B=1$, $\mu_i=2$, $J=\Delta=1$, $\beta_h/\beta_c=0.2$, and $L=2000$.
  All quantities are dimensionless.}
  \label{WQck}
 \end{figure}

According to the first law of thermodynamics, the net work extracted by the quantum Stirling cycle is
\begin{align}
   W_S&=Q_\mathrm{I}+Q_\mathrm{II}+Q_\mathrm{III}+Q_\mathrm{IV} \notag \\
   &=\sum_{k>0}
    \left\{\frac{2}{\beta_h}\left[\ln\cosh\left(\frac{\beta_h\varepsilon_k^f}{2}\right)-\ln\cosh\left(\frac{\beta_h
          \varepsilon_k^i}{2}\right)\right]\right. \notag \\
          &+\left.\frac{2}{\beta_c}\left[\ln\cosh\left(\frac{\beta_c\varepsilon_k^i}{2}\right)-\ln\cosh\left(\frac{\beta_c
          \varepsilon_k^f}{2}\right)\right]\right\}.
\end{align}
For each cycle, the amount of heat $Q_S^h$ absorbed by the working medium reads
\begin{align}
  Q_S^h&=Q_\mathrm{I}+Q_\mathrm{IV} \notag \\
      &=\sum_{k>0}
      \left\{\frac{2}{\beta_h}\left[\ln\cosh\left(\frac{\beta_h\varepsilon_k^f}{2}\right)-\ln\cosh\left(\frac{\beta_h
          \varepsilon_k^i}{2}\right)\right]\right. \notag \\
         &+\left.\left[\varepsilon_k^i\tanh\left(\frac{\beta_c\varepsilon_k^i}{2}\right)-\varepsilon_k^f\tanh\left(
          \frac{\beta_h\varepsilon_k^f}{2}\right)\right]\right\}.
\end{align} 
Therefore the efficiency of the heat engine reads
\be 
  \eta_S=\frac{W_S}{Q_S^h}.
\ee
We stress that in order to extract work from the engine we should have $W_S>0$, as in the Otto cycle,
and the efficiency $\eta_S$ is also limited by the Carnot efficiency $\eta_c$ \cite{MaY2017}.

Based on the above results, in the next section we investigate the effects of the long-range
interactions on the performance of a quantum heat engine which drives through 
the quantum Otto and Stirling cycles, respectively.

\section{Results and discussions} \label{NRD}

In our numerical calculation, the chemical potential varies from the initial value $\mu_i=2$ to the final value $\mu_f$
with $0\leq\mu_f\leq2$.
To unveil the influences of the long-range interactions on the performance of the quantum engine, 
for both Otto and Stirling cycles we consider the following ratios:
\be \label{DfR}
   \mathcal{R}^{\kappa}_W=\frac{W_{\kappa}}{W_{\kappa}^\infty},\ 
   \mathcal{R}^{\kappa}_\eta=\frac{\eta_{\kappa}}{\eta_{\kappa}^\infty},
\ee
where $\kappa=O$ ($S$) corresponds to the Otto (Stirling) cycle, 
$W_{\kappa}$ ($\eta_{\kappa}$)  is the work output (efficiency) 
with finite $\alpha$, while $W_{\kappa}^\infty$ ($\eta_{\kappa}^\infty$) represents its 
short-range (with $\alpha=\infty$) counterpart. 

  \begin{figure}
    \includegraphics[width=\columnwidth]{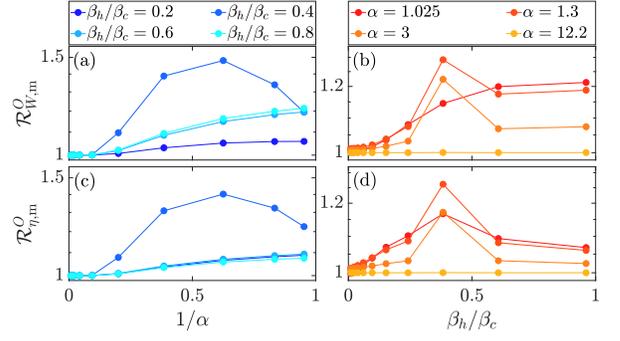}
  \caption{Left column: The maximum work output ratio $\mathcal{R}_{W,\mathrm{m}}^O$ (a) 
  and the maximum efficiency ratio $\mathcal{R}_{\eta,\mathrm{m}}^O$ (c) of Otto cycle 
  as a function of $\alpha$, with different curves in each panel representing different ratios of bath temperatures.
  Right column: $\mathcal{R}_{W,\mathrm{m}}^O$ (b) and
  $\mathcal{R}_{\eta,\mathrm{m}}^O$ (d) as a function of $\beta_h/\beta_c$ for different $\alpha$ (different colors).
  Parameters are $\hbar=k_B=1$, $\beta_c=5$, $J=\Delta=1$, and $L=2000$.}
  \label{MaxRwa}
 \end{figure}

The ratios defined in Eq.~(\ref{DfR}) compare the net work output and efficiency of 
a long-range interacting heat engine with that of a short-range interacting heat engine. 
If $\mathcal{R}_W^\kappa=1$ and $\mathcal{R}_\eta^\kappa=1$, the long-range interacting heat engine has
the same performance as its short-range counterpart.
Conversely, having $\mathcal{R}_W^\kappa>1, \mathcal{R}_\eta^\kappa>1$ indicates that there is an enhancement in  
engine performance induced by the long-range interactions. 
Finally, $\mathcal{R}_W^\kappa<1, \mathcal{R}_\eta^\kappa<1$ mean that the quantum heat engine with short-range
interacting working medium is more beneficial.
In our following study, we will investigate the dependence of these ratios 
on the interacting range $\alpha$ for the Otto and Stirling cycles, respectively.

We separately consider the cases in which the quantum engine operating at low and high temperature 
with the ratio of bath inverse temperature $\beta_h/\beta_c$ is fixed.
For the low-temperature case, we take the inverse temperature of the cold bath as $\beta_c=5$, while for the 
high-temperature case we chose $\beta_c=0.05$. 
As we have pointed out in Sec.~\ref{Model}, we focus on the situation $\alpha>1$, 
where a transition between the long and short ranges occurs. 

Before we present our results for a specific thermodynamic cycle, we first illustrate the notable differences
in the behavior of quasiparticle energy $\varepsilon_k$ [cf.~Eq.~(\ref{quasiE})] arising from the long-range interactions.  
Figure~\ref{EngV} depicts the quasiparticle energy as a function of momentum $k$ and $\mu_f$ for different $\alpha$. 
By comparing with the short-range interactions case [panel (d)], one can see that the 
long-range interactions have strong impacts on the behavior of quasiparticle energy. 
As the work output and efficiency of a heat engine involve the sum of quasiparticle energy, the differences in the 
quasiparticle energy from the interacting range allow us to expect that
long-range interactions should affect the performance of the quantum engine.  

 \begin{figure}
    \includegraphics[width=\columnwidth]{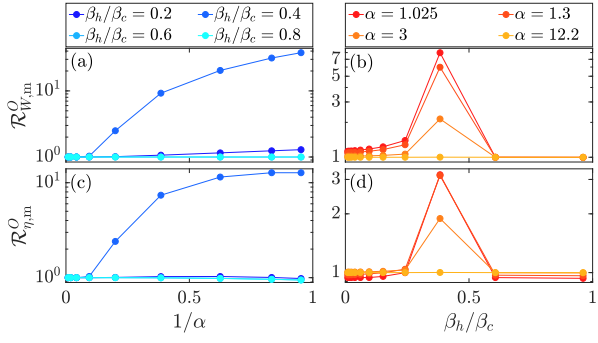}
  \caption{Same as Fig.~\ref{MaxRwa} but for the
   high-temperature case with $\beta_c=0.05$.}
  \label{MaxRwb}
 \end{figure}

\subsection{Results for quantum Otto cycle}

By using the results outlined in Sec.~\ref{QOC}, we are able to calculate the work output $W_O$ and efficiency $\eta_O$ 
of the Otto cycle for different interacting ranges and cycle parameters.
We first focus on the behavior of work output ratio $\mathcal{R}_W^O$ for several ranges of interactions.

In the first row of Fig.~\ref{OtRv}, we plot $\mathcal{R}_W^O$ as a function of $\mu_f/\mu_i$ 
for different values of $\alpha$ at 
$\beta_c=5$ [panel (a)] and $\beta_c=0.05$ [panel (b)].
We first notice that for some chemical potential regimes the value of $\mathcal{R}_W^O$ 
can be greater than $1$ and increases with
decreasing $\alpha$.  
This means that the long-range interactions in the Kitaev chain can enhance the performance of the Otto engine.
Specifically, for the engine operating at low temperature with $\beta_c=5$, the work output will get 
a significant enhancement with increasing (decreasing) the range of interactions ($\alpha$)
when $\mu_f/\mu_i>0.5$, as shown in Fig.~\ref{OtRv}(a).
On the contrary, the work output is enhanced by the long-range interactions in the region $\mu_f/\mu_i<0.5$ for the 
high temperature case with $\beta_c=0.05$, as evidenced by Fig.~\ref{OtRv}(b).

 \begin{figure*}
  \includegraphics[width=\textwidth]{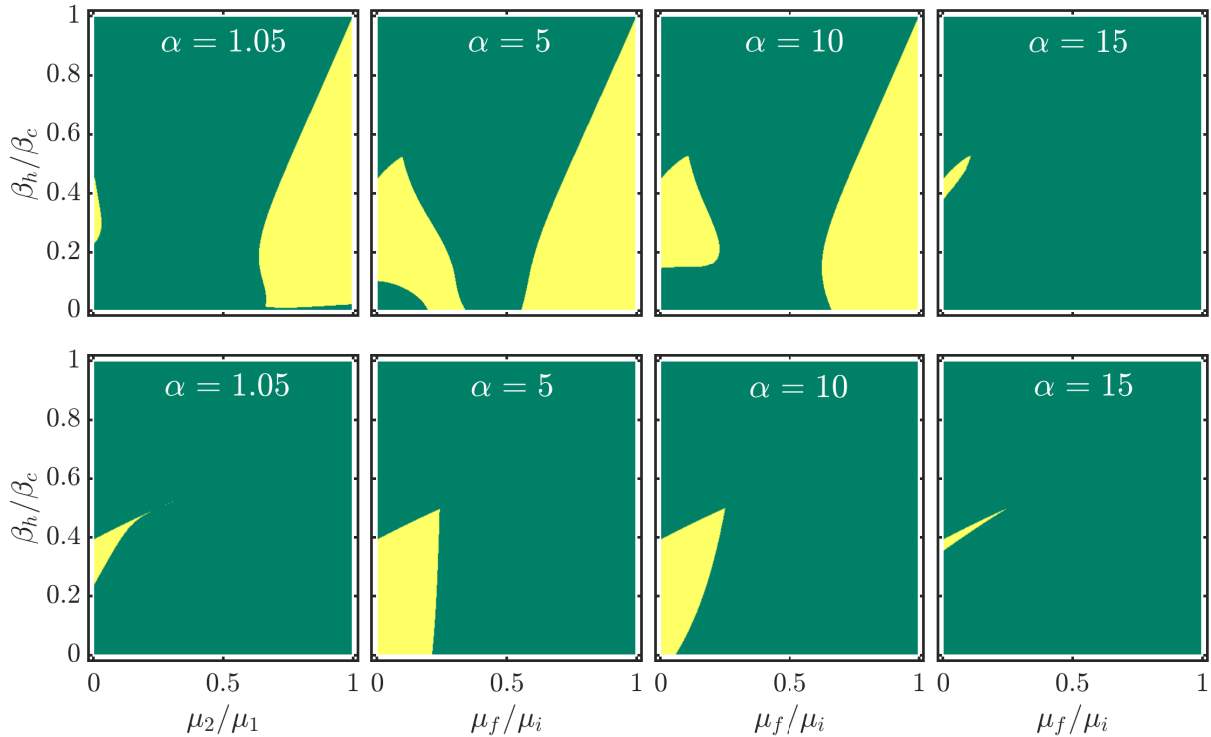}
  \caption{Enhancement regions of the Otto cycle in parameter space for different values of $\alpha$. 
  Here the enhancement regions are identified as the regions with $\mathcal{R}_W^O>1$ and $\mathcal{R}_\eta^O>1$.
  The yellow regions indicate the enhancement regions in parameter space under which
  the performance of the heat engine is enhanced by long-range interactions.
  The upper panels are the results for the low-temperature case with $\beta_c=5$, and the bottom panels represent the 
  high-temperature case with $\beta_c=0.05$. The remaining parameters are $\hbar=k_B=1$, $\mu_i=2$, $J=\Delta=1$,
  and $L=2000$.}
  \label{RegEn}
 \end{figure*}

We further see that $\mathcal{R}_W^O$ is minimized at around  $\mu_f/\mu_i=0.5$, 
which corresponds to the critical point of the system. 
In particular, for the case of low temperature, $\mathcal{R}_W^O$ shows a remarkable decrease at
long-range interactions.
It is known that the work output takes its local minimal value at the critical point of the system \cite{Fadaie2018}. 
The dramatic drop in $\mathcal{R}_W^O$ near $\mu_f/\mu_i=0.5$ implies that the negative impact
of the phase transition on the extractable work can be enhanced by long-range interactions.

The efficiency ratio $\mathcal{R}_\eta^O$ for different ranges of interactions at low and high temperature cases
are displayed in the second row of Fig.~\ref{OtRv}.
As shown in Fig.~\ref{OtRv}(c), at low temperature with $\beta_c=5$, the efficiency for a long-range interacting engine
is always superior to its short-range counterpart as long as $\mu_f/\mu_i\neq0.5$.
Around the critical point $\mu_f/\mu_i=0.5$, $\mathcal{R}_\eta^O$ also has a significant decrease 
for long-range interactions, indicating the enhanced negative impact on the efficiency at the critical point.

For the case of high temperature with $\beta_c=0.05$,  as can be seen from Fig.~\ref{OtRv}(d),
the enhancement in efficiency due to the long-range interactions occurs only for 
smaller $\mu_f/\mu_i$ when $\alpha\geq1.2$. 
For $1<\alpha<1.2$, the long-range interactions show a universal negative impact on the efficiency of the engine.   
Moreover, we still see $\mathcal{R}_\eta^O$ is minimized around the critical point in the case of high temperature.
We stress that the efficiency is always less than Carnot efficiency, 
regardless of interacting ranges and bath temperatures.

The features observed for $\mathcal{R}_W^O$ and $\mathcal{R}_\eta^O$ can be understood as follows.
We first rewrite $\mathcal{R}_W^O$ and $\mathcal{R}_\eta^O$ as
\begin{align} \label{RwRe}
   &\mathcal{R}_W^O=1-\xi_O\frac{\Delta Q_O^h}{Q_O^{h,\infty}},  \notag \\
   &\mathcal{R}_\eta^O=1-\frac{\Delta Q_O^h}{Q_O^{h,\infty}}\frac{\xi_O-1}{1-\Delta Q_O^h/Q_O^{h,\infty}}, 
\end{align}
where $\Delta Q_O^h=Q_O^{h,\infty}-Q_O^h$ is the absorbed heat
differences between short- and long-range engines, and $\xi_O=\Delta W_O/(\eta_O^\infty\Delta Q_O^h)$
with $\Delta W_O=W_O^\infty-W_O$.
We thus show that the $\Delta Q_O^h/Q_O^{h,\infty}$ and $\xi_O$ act as the indicators of enhancement in
engine performance.
Figure~\ref{WQck} displays $\Delta Q_O^h/Q_O^{h,\infty}$ and $\xi_O$ versus $\alpha$ for
several representative values of $\mu_f/\mu_i$ with $\beta_c=5$ [column (a)] and $\beta_c=0.05$ [column (b)].

At low temperature [column (a) in Fig.~\ref{WQck}], as can be seen from the upper panel, 
$\Delta Q_O^h/Q_O^{h,\infty}$ is always greater than zero and decreases with increase in $\alpha$, 
irrespective of the value of $\mu_f/\mu_i$.
On the other hand, for all ranges of interactions studied, the values of $\xi_O$ are less than zero  
in the region $\mu_f/\mu_i>0.5$.
According to Eq.~(\ref{RwRe}), this makes both $\mathcal{R}_W^O$ and $\mathcal{R}_\eta^O$ 
greater than $1$ for $\mu_f/\mu_i>0.5$. 
For the case of $\mu_f/\mu_i<0.5$, we notice that $0<\xi_O<1$ when $\alpha<2$, while as 
$\alpha>2$, the value of $\xi_O$ becomes less than zero (see the inset in the bottom panel).
This indicates that in the region $\mu_f/\mu_i<0.5$, $\mathcal{R}_W^O<1$ for the system with long-range interactions
and increases with decreasing the range of interactions.
However, as $\xi_O<1$ holds for any value of $\alpha$, $\mathcal{R}_\eta^O$ will always be
greater than $1$ at $\mu_f/\mu_i<0.5$.
At the critical point $\mu_f/\mu_i=0.5$, we find that $\Delta Q_O^h/Q_O^{h,\infty}$ is maximized, 
regardless of the values of $\alpha$. 
Meanwhile, the behavior of $\xi_O$ shows that $\xi_O>1$ at smaller $\alpha$ and it approximately equal to $1$ 
for $\alpha>2$. 
These two factors give rise to the remarkable decrease in $\mathcal{R}_W^O$ and $\mathcal{R}_\eta^O$
at $\mu_f/\mu_i=0.5$.

In column (b) of Fig.~\ref{WQck} we demonstrate the same analysis for the case of high temperature.
For all the ranges of interactions studied, $\Delta Q_O^h/Q_O^{h,\infty}<0$ and $\xi_O>0$ at $\mu_f/\mu_i<0.5$.
This leads to $\mathcal{R}_W^O>1$ for small values of the ratio $\mu_f/\mu_i$.
In addition, as shown in the inset of the bottom panel, for $\mu_f/\mu_i<0.5$, at smaller $\alpha$ 
we have $\xi_O<1$, but for greater values of $\alpha$, we have $\xi_O>1$. 
This means that the efficiency ratio changes from $\mathcal{R}_\eta^O<1$ for long-range interactions to
$\mathcal{R}_\eta^O>1$ as the range of interactions decreases in the region $\mu_f/\mu_i<0.5$.
However, when $\mu_f/\mu_i\geq0.5$, we see that regardless of the range of interactions, 
the second term $\xi_O\Delta Q_O^h/Q_O^{h,\infty}$ in $\mathcal{R}_W^O$ [cf.~Eq.~(\ref{RwRe})] 
is always greater than zero, despite that $\Delta Q_O^h/Q_O^{h,\infty}$ can vary from 
negative values to positive values (see the inset in the upper panel).
As a consequence, the value of $\mathcal{R}_W^O$ is less than $1$ for greater values of $\mu_f/\mu_i$. 
Moreover, as $\Delta Q_O^h/Q_O^{h,\infty}<0 (>0)$ is associated with $\xi_O<0 (>1)$ 
at $\mu_f/\mu_i\geq0.5$, the value of $\mathcal{R}_\eta^O$ is also less than $1$. 
Finally, in the high-temperature case, the maximal value of $|\Delta Q_O^h/Q_O^{h,\infty}|$ 
is still reached at the critical point,
indicating $\mathcal{R}_W^O$ and $\mathcal{R}_\eta^O$ have a significant decrease at $\mu_f/\mu_i=0.5$.

\begin{figure}
  \includegraphics[width=\columnwidth]{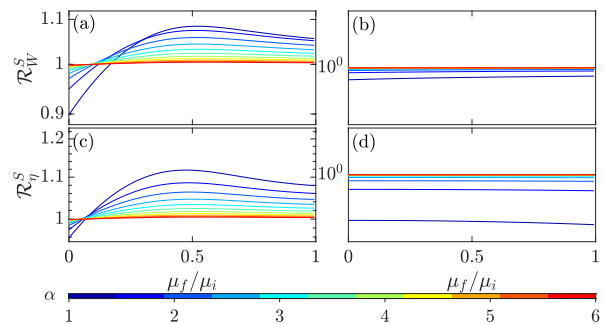}
  \caption{Work output ratio $\mathcal{R}_W^S$ of the Stirling cycle as a function of 
  $\mu_f/\mu_i$ with $\beta_c=5$ (a) and $\beta_c=0.05$ (b) and
  the efficiency ratio $\mathcal{R}_\eta^S$ as a function of $\mu_f/\mu_i$ 
  with $\beta_c=5$ (c) and $\beta_c=0.05$ (d) for $\alpha\in[1.05,6]$.
  The remaining parameters are $\hbar=k_B=1$, $\mu_i=2$, $\beta_h/\beta_c=0.2$, $J=\Delta=1$, and $L=2000$.}
  \label{RwSt}
\end{figure}

To gain further insight into the details of the long-range interaction impacts on the performance of the quantum engine, 
we plot $\mathcal{R}_{W,\mathrm{m}}^O$ and $\mathcal{R}_{\eta,\mathrm{m}}^O$, 
defined as the maximum in $\mathcal{R}_W^O$ and $\mathcal{R}_\eta^O$,
versus $\alpha$ for different values of $\beta_h/\beta_c$ with $\beta_c=5$
in the left column of Fig.~\ref{MaxRwa}.
On can immediately see a striking similarity between $\mathcal{R}_{W,\mathrm{m}}^O$ and
$\mathcal{R}_{\eta,\mathrm{m}}^O$.
Regardless of the value of $\beta_h/\beta_c$, both of them converge towards unity 
in the short-range limit $\alpha\to\infty$, as expected.
However, at $\beta_h/\beta_c=0.4$, we see that 
$\mathcal{R}_{W,\mathrm{m}}^O$ and $\mathcal{R}_{\eta,\mathrm{m}}^O$ exhibit a remarkably 
different behavior as compared to the results obtained for other values of $\beta_h/\beta_c$. 
Specifically, at $\beta_h/\beta_c=0.4$, both $\mathcal{R}_{W,\mathrm{m}}^O$ and 
$\mathcal{R}_{\eta,\mathrm{m}}^O$ first experience a growth and then decrease to unity as $\alpha$ increases, 
whereas we observe a monotonous decrement in $\mathcal{R}_{W,\mathrm{m}}^O$ and
$\mathcal{R}_{\eta,\mathrm{m}}^O$ with increase in $\alpha$ for $\beta_h/\beta_c\neq0.4$.
Moreover, the observed $\beta_h/\beta_c$ dependence in the behaviors of $\mathcal{R}_{W,\mathrm{m}}^O$ 
and $\mathcal{R}_{\eta,\mathrm{m}}^O$ imply that the bath temperatures can strongly influence 
the enhancing ability of long-range interactions.   
In the right column of Fig.~\ref{MaxRwa}, we show $\beta_h/\beta_c$ dependence of $\mathcal{R}_{W,\mathrm{m}}^O$
and $\mathcal{R}_{\eta,\mathrm{m}}^O$ for several values of $\alpha$.
We see again the remarkable resemblance between $\mathcal{R}_{W,\mathrm{m}}^O$ and 
$\mathcal{R}_{\eta,\mathrm{m}}^O$ when $\alpha>1.025$.
At $\alpha=1.025$, we can see that $\mathcal{R}_{W,\mathrm{m}}^O$ increases
monotonously as $\beta_h/\beta_c$ increases, 
while $\mathcal{R}_{\eta,\mathrm{m}}^O$ shows a growth followed by a decrement with increase in $\beta_h/\beta_c$.
The features in Fig.~\ref{MaxRwa} indicate that for an Otto cycle operating at low temperature, 
there exists an optimal condition for which the maximum enhancement in the work output and
efficiency induced by long-range interactions are reached at the same time.  
For the case considered here, it is approximately given by $\alpha\approx1.5$, $\beta_h/\beta_c\approx0.4$.

 \begin{figure}
  \includegraphics[width=\columnwidth]{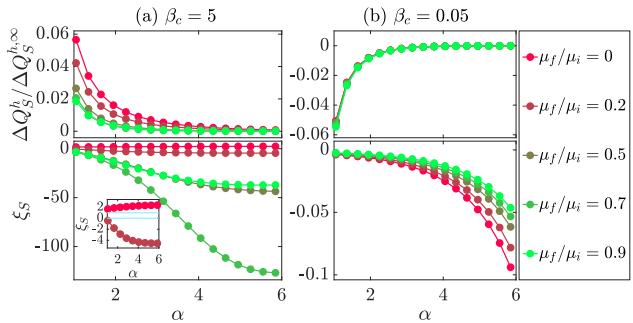}
  \caption{Column (a): $\Delta Q_S^h/Q_S^{h,\infty}$ (upper panel) and 
  $\xi_S$ (bottom panel and zoomed-in inset) as a function of $\alpha$ for different values of 
  $\mu_f/\mu_i$ at $\beta_c=5$. 
  The solid horizontal lines in the inset indicate $\xi_S=0$ and $\xi_S=1$, respectively.
  Column (b): The same as in column (a) applied to the high-temperature case with $\beta_c=0.05$. 
  The remaining parameters are $\hbar=k_B=1$, $\mu_i=2$, $\beta_h/\beta_c=0.2$, $J=\Delta=1$, and $L=2000$.
  All quantities are unitless.}
  \label{DqW}
 \end{figure}

For the high temperature case, the $\alpha$ and $\beta_h/\beta_c$ dependencies of $\mathcal{R}_{W,\mathrm{m}}^O$ 
and $\mathcal{R}_{\eta,\mathrm{m}}^O$ are shown in Fig.~\ref{MaxRwb}.
As can be seen from the left column of Fig.~\ref{MaxRwb},
the behavior of $\mathcal{R}_{W,\mathrm{m}}^O$ is qualitatively 
similar to $\mathcal{R}_{\eta,\mathrm{m}}^O$.
Namely, both of them are decreased with an increase in $\alpha$ and approaching unity as $\alpha\to\infty$, 
irrespective of the value of $\beta_h/\beta_c$.
On the other hand, for different values of $\alpha$, the observed $\beta_h/\beta_c$ dependence of 
$\mathcal{R}_{W,\mathrm{m}}^O$ is also very similar to $\mathcal{R}_{\eta,\mathrm{m}}^O$, 
as evidenced by the right column in Fig.~\ref{MaxRwb}.
The results in Fig.~\ref{MaxRwb} further verify that the above-defined optimal condition also existed    
in the case of high temperature. 
The optimal condition for the case studied here is given by $\alpha=1.025$, $\beta_h/\beta_c=0.4$, 
different from the case of low temperature.
By comparing Figs.~\ref{MaxRwa} and \ref{MaxRwb}, we further notice that there is a greater enhancement 
in engine performance at the high temperature.

It is also of interest to study how the regions of enhancement in parameter space evolve as the 
range of interactions is changed from the long-rang to short-range limit. 
In Fig.~\ref{RegEn} we plot the regions where the enhancement
is absent (dark green) or present (yellow) for several values of 
$\alpha$ with $\beta_c=5$ (upper row) and $\beta_c=0.05$ (bottom row).
Here, in our study, the enhancement regions are the regions with $\mathcal{R}_W^O>1$ and $\mathcal{R}_\eta^O>1$.  
We see that for both the low- and high-temperature cases, the enhancement regions first undergo an expansion in 
parameter space and then shrink with increasing $\alpha$. 
One can expect that such regions will vanish as $\alpha\to\infty$.
Here, it is worth mentioning that $\mathcal{R}_W^O$ and $\mathcal{R}_\eta^O$ 
decrease with increasing $\alpha$ and approach $1$ as $\alpha\to\infty$.
The expansion of the enhancement regions is, therefore, associated with the reduction 
in the enhancing ability of the long-range interactions.
We finally note that the size of the enhancement regions in the case of low temperature is always 
lager than the corresponding high-temperature case.

 \begin{figure}
  \includegraphics[width=\columnwidth]{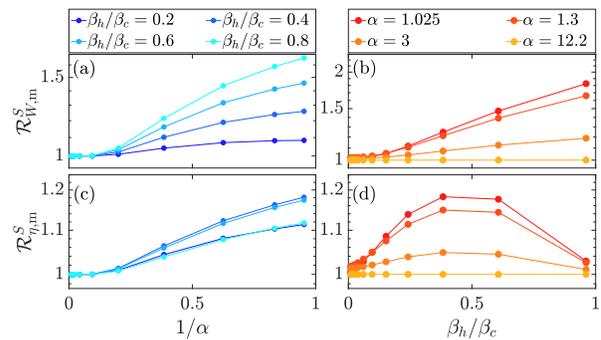}
  \caption{Left column:
  The maximum work output ratio $\mathcal{R}_{W,\mathrm{m}}^S$ (a) and
  the maximum efficiency ratio $\mathcal{R}_{\eta,\mathrm{m}}^S$ (c) of the Stirling cycle as a function of $\alpha$
  for different values of $\beta_h/\beta_c$.
  Right column: $\mathcal{R}_{W,\mathrm{m}}^S$ (b) and 
  $\mathcal{R}_{\eta,\mathrm{m}}^S$ (d) as a function of $\beta_h/\beta_c$ for several $\alpha$. 
  The remaining parameters are $\hbar=k_B=1$, $\beta_c=5$, $J=\Delta=1$, 
  and $L=2000$.}
  \label{MaxRSt}
 \end{figure}

\subsection{Results for quantum Stirling cycle}

We continue our study by exploring the long-range interaction effects on the performance of the quantum Stirling cycle.
According to the results in Sec.~\ref{QSC}, we have calculated the work output ratio $\mathcal{R}_W^S$
and efficiency ratio $\mathcal{R}_\eta^S$ defined in
Eq.~(\ref{DfR}) for a Stirling cycle with different ranges of interactions and cycle parameters; the results are shown in
Fig.~\ref{RwSt}.

First, for the case of low temperature with $\beta_c=5$, both the work output and efficiency of 
the Stirling cycle are enhanced
by the long-range interactions at greater $\mu_f/\mu_i$; see panels (a) and (c) in Fig.~\ref{RwSt}. 
Moreover, different from the results in the Otto cycle, we see that the maximal values of $\mathcal{R}_W^S$ 
and $\mathcal{R}_\eta^S$ are obtained near the critical point $\mu_f/\mu_i=0.5$, 
and increase with decreasing $\alpha$.
Hence, for the Stirling cycle, the enhancement effect of long-range interactions can be boosted by 
the critical point. 
On the contrary, at high temperature with $\beta_c=0.05$, as can be seen from 
panels (b) and (d) in Fig.~\ref{RwSt}, we have $\mathcal{R}_W^S<1$ and $\mathcal{R}_\eta^S<1$ 
over the whole parameter region. 
Therefore, the long-range interactions are useless to improve the performance of the Stirling cycle if it is 
operating at high temperature.

 \begin{figure}
  \includegraphics[width=\columnwidth]{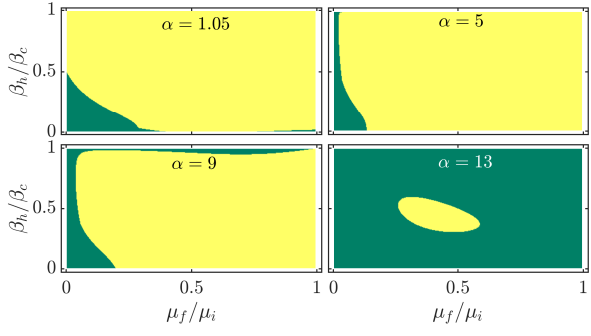}
  \caption{Regions of enhancement in the parameter space of a Stirling cycle for different values of $\alpha$.
  The enhancement regions are defined as the regions with $\mathcal{R}_W^S>1$ and $\mathcal{R}_\eta^S>1$. 
  Yellow and dark green shaded domains indicate the regions with enhancement and no enhancement.
  The parameters are: $\hbar=k_B=1$, $\mu_i=2$, $\beta_c=5$, $\beta_h/\beta_c=0.2$, $J=\Delta=1$, 
  and $L=2000$.}
  \label{StER}
 \end{figure}

To understand the features observed in Fig.~\ref{RwSt}, as we have done for the case of the Otto cycle, we recast 
$\mathcal{R}_W^S$ and $\mathcal{R}_\eta^S$ as follows:
\begin{align}
   &\mathcal{R}_W^S=1-\xi_S\frac{\Delta Q_S^h}{Q_S^{h,\infty}}, \notag \\
   &\mathcal{R}_\eta^S=1-\frac{\Delta Q_S^h}{Q_S^{h,\infty}}\frac{\xi_S-1}{1-\Delta Q_S^h/Q_S^{h,\infty}},
\end{align}
where $\Delta Q_S^h=Q_S^{h,\infty}-Q_S^h$ is the absorbed heat
difference of the Stirling cycle, and $\xi_S=\Delta W_S/(\eta_S^\infty\Delta Q_S^h)$
with $\Delta W_S=W_S^\infty-W_S$.
Then, we consider the dependences of $\Delta Q_S^h/Q_S^{h,\infty}$ and $\xi_S$ on the range of interactions 
for different values of $\mu_f/\mu_i$.
Column (a) of Fig.~\ref{DqW} shows $\Delta Q_S^h/Q_S^{h,\infty}$ (upper panel) and $\xi_S$ (bottom panel) 
versus $\alpha$ for a few representative values of $\mu_f/\mu_i$ at low temperature. 
Clearly, we always have $\Delta Q_S^h/Q_S^{h,\infty}>0$ for all ranges of interactions and values of $\mu_f/\mu_i$.
Therefore there are negative values of $\xi_S$ at $\mu_f/\mu_i\geq0.2$ (see the bottom panel), indicating 
$\mathcal{R}_S^W>1$ and $\mathcal{R}_\eta^S>1$.
For the case of $\mu_f/\mu_i=0$, as can be seen from the inset in the bottom panel, we always have $\xi_S>1$,
resulting in $\mathcal{R}_W^S<1$ and $\mathcal{R}_\eta^S<1$.
In marked contrast to the low-temperature case, at high temperature with $\beta_c=0.05$ 
we find that $\Delta Q_S^h/Q_S^{h,\infty}<0$ and $\xi_S<0$, 
regardless of the values of $\alpha$ and $\mu_f/\mu_i$, as shown in column (b) of Fig.~\ref{DqW}.
We thus have $\mathcal{R}_W^S<1$ and $\mathcal{R}_\eta^S<1$, independent of $\alpha$ and $\mu_f/\mu_i$.

As we focus on the enhancement effect of the long-range interactions, we only consider the 
low-temperature case in the following studies of the Stirling cycle.
Figure~\ref{MaxRSt} depicts the maximum work output ratio $\mathcal{R}_{W,\mathrm{m}}^S$ and the maximum efficiency
ratio $\mathcal{R}_{\eta,\mathrm{m}}^S$ in the Stirling cycle.
We see that as $\alpha$ increases, $\mathcal{R}_{W,\mathrm{m}}^S$ and $\mathcal{R}^S_{\eta,\mathrm{m}}$ 
converge towards $1$ in a similar way; see panels (a) and (c) in Fig.~\ref{MaxRSt}. 
However, the dependence of $\mathcal{R}_{W,\mathrm{m}}^S$ on $\beta_h/\beta_c$ is prominent, different from
the case of maximum efficiency ratio, as visible in panels (b) and (d) of Fig.~\ref{MaxRSt}.
Specifically, $\mathcal{R}_{W,\mathrm{m}}^S$ increases monotonously with an increase in 
$\beta_h/\beta_c$, while the increment in $\mathcal{R}_{\eta,\mathrm{m}}^S$ at smaller $\beta_h/\beta_c$ is  
followed by a decrement at greater $\beta_h/\beta_c$.
As a consequence, the optimal condition extracted from 
$\mathcal{R}_{W,\mathrm{m}}^S$ in Figs.~\ref{MaxRSt}(a) and \ref{MaxRSt}(b) is inconsistent with 
that obtained from the results of $\mathcal{R}_{\eta,\mathrm{m}}^S$ 
[see Figs.~\ref{MaxRSt}(c) and \ref{MaxRSt}(b)].
This disagreement means that the maximum enhancement in work output and efficiency 
cannot be achieved simultaneously by the long-range interactions in the Stirling cycle. 
This is in sharp contrast to the Otto cycle, where the
work output ratio and efficiency ratio can be maximized under the same condition.

We finally investigate the evolution of enhancement regions in the parameter space of the Stirling cycle with decreasing
the range of interactions. 
As in the Otto cycle, the enhancement regions in the Stirling cycle are still identified as the regions with
$\mathcal{R}_W^S>1$ and $\mathcal{R}_\eta^S>1$.
Figure~\ref{StER} shows the enhancement regions of the Stirling cycle for several values of $\alpha$.
We see that as $\alpha$ increases, the enhancement regions first experience an expansion and then shrink in the
parameter space, similar to the case of the Otto cycle.  
As expected, the enhancement regions will disappear in the short-range limit $\alpha\to\infty$.
Moreover, such as in the Otto cycle, due to the values of $\mathcal{R}_{W}^S$ and 
$\mathcal{R}_{\eta}^S$ decreasing with increasing $\alpha$,
the expansion of the enhancement regions will associate with a larger reduction 
in the enhancing ability of the long-range interactions.

\section{Conclusions} \label{ConS}

In conclusion, we have examined how the long-range interactions in quantum many-body systems affect the performance
of quantum heat engines.
To this end, we have introduced a many-body quantum heat engine with a long-range Kitaev chain as its working medium.
The integrability nature of the long-range interacting Kitaev chain allows us to study the performance of 
the heat engine through fully analytical results.
We have demonstrated that the long-range interactions exhibit remarkable and nontrivial impacts on 
the performance of the quantum heat engine.
We have considered quantum heat engine operators
with different thermodynamic cycles--the Otto cycle and the Stirling cycle.

For the Otto cycle operating at low temperature, we found that in all studied figures of merit of engine performance, 
including net work output and efficiency, the long-range interactions 
enhance the engine performance for certain cycle parameter regimes. 
This enhancement effect still persists when the engine is working at high temperature. 
However, near the critical point of the system, the work output and efficiency decrease dramatically with 
increasing the range of interactions, irrespective of the bath temperatures.
In the case of a Stirling cycle at low temperature, the long-range interactions exhibit a similar enhancement effect on
the performance of the engine.
However, different from the Otto cycle, in this case the long-range interactions lead to a maximum enhancement 
of the performance of the Stirling cycle around the critical point of the system.
We found the universal negative impact of the long-range interactions when 
the Stirling cycle operated at high temperature.

For both cycles we investigated the possibility of long-range interactions leading to a
maximum enhancement in work output and efficiency of the quantum heat engine at the same time.
We found that in the Otto cycle the maximum enhancement in the performance of 
the engine can be achieved simultaneously by
the long-range interactions, while this is not true in the case of the Stirling cycle.
In addition, we also explored the dependencies of the enhancement regions in the cycle parameter space on the range 
of interactions and showed that with a decrease in range of interactions the enhancement regions 
first expand and then quickly shrink.
Such regions finally vanish as $\alpha\to\infty$.  

Our results evidence that the long-range interactions in many-body systems have strong effects on the performance
of quantum heat engines. 
Therefore, our work provides an additional insight into how to improve engine performance 
in quantum many-body heat engines \cite{Jaramillo2016,Beau2016,LiT2018,Chen2019,Halpern2019,
Kloc2019,Hartmann2019,Revathy2020}.  
Moreover, the experimental realization of one-dimensional long-range Kitaev chains in physical systems has been 
proposed in many studies \cite{TongQ2013,Benito2014,Pientka2013,Pientka2014,Giuliano2018}.
In particular, by using an experimental platform based on a superconductor in proximity
to a two-dimensional electron gas with strong spin-orbit coupling \cite{Hell2017,Pientka2017},    
a very recent work has been demonstrated that long-range Kitaev chains can be obtained
via planar Josephson junctions \cite{Dillon2018}.
In this scheme, tuning of the range of interactions can be achieved by controlling the strength of the Zeeman field,
whereas the tunable chemical potential is accomplished through the phase difference between two $s$-wave
superconductors.
On the other hand, several proposals for realizing quantum heat engines in superconducting devices 
have been discussed in a variety of works \cite{Campisi2015,Marchegiani2016}.
These facts lead us to believe that our results could be experimentally verified.

We expect the performance enhancement of quantum engines by long-range interactions 
still holds in other long-range interacting systems, such as long-range interacting spin systems. 
It will be interesting to systematically study how the range of interactions affect the engine performance 
in different long-range interacting systems. 
Another interesting extension of the present work is to analyze the role of the range of interactions
in finite time cycles.  
By considering that long-range interacting spin systems have been realized in recent experiments
\cite{Jurcevic2014,Richerme2014},  
we hope that our present results would be able to trigger more experimental efforts to investigate the effects
of long-range interactions on the performance of quantum heat engines.

\acknowledgements 

Q.~W. acknowledges support from the National Science Foundation of China under Grant No.~11805165, 
Zhejiang Provincial Nature Science Foundation under Grant No.~LY20A050001, 
and the Slovenian Research Agency (ARRS) under the Grants
No.~J1-9112 and No.~P1-0306.  

\bibliographystyle{apsrev4-1}
\bibliography{Draft1}

\end{document}